\documentclass[proceedings, preprint]{rmaa}

\newcommand{\D}{\discretionary{}{}{}}

\SetYear{2010}
\SetConfTitle{XIII Latin American Regional IAU Meeting (2010), M\'exico.}

\title{Spectroscopy of BL Cam - observations}    

\author{ M. Alvarez\altaffilmark{1}, R. Michel\altaffilmark{1}, L. Fox-Machado\altaffilmark{1},  
JP. Sareyan\altaffilmark{2}, S. Fauvaud\altaffilmark{3}, G. Tovmassian\altaffilmark{1} } 

\altaffiltext{1}{OAN, Instituto de  Astronom\'\i{}a, UNAM, Ensenada, BC, 22860,  M\'exico, (alvarez\D{}@astrosen.\D{}unam.\D{}mx).} 
\altaffiltext{2}{Observatoire de la C$\hat{o}$te dÕAzur, Nice, France,  (jean-pierre.sareyan\D{}@obspm.\D{}fr).} 
\altaffiltext{3}{Observatoire  du Bois de Bardon, 16110 Taponnat, France, (stephane.fauvaud\D{}@wanadoo.\D{}fr).} 
\suppressfulladdresses

\shortauthor{Alvarez, Michel, Fox-Machado et al.}
\shorttitle{Spectroscopy of BL Cam }

\listofauthors{M. \'Alvarez,  \ R. Michel, \ L. Fox-Machado,
 \ JP. Sareyan, \ S. Fauvaud, \ G. Tovmassian } 

\indexauthor{\'Alvarez, M.}
\indexauthor{Michel, R.}
\indexauthor{Fox-Machado, L.}
\indexauthor{Sareyan, JP.}
\indexauthor{Fauvaud, S.}
\indexauthor{Tovmassian, G.}

\addkeyword{Stars: individual: BL Camelopardalis}
\addkeyword{Stars: individual: GD 428}
\addkeyword{Stars: individual: SX Phe stars}
\addkeyword{Stars: pulsating variable stars}
\addkeyword{techniques: spectroscopy}
\addkeyword{techniques: CCD Johnson's and Stromgren's photometry}

\begin{document}

\maketitle

 \section{Recent Results}


The star BL Camelopardalis (GD 428, GSC 04067-00471, V=12.92, B=13.1), is a SX Phoenicis pulsating variable showing a high amplitude variability (300-350 mmag), with a very clear period of 56.3 min, that has being recently studied by Fauvaud et al. (2010).  They show that the observed variability is due to a clear and single pulsation, as expected for these type of pulsators. It was discovered as a variable star on 1976 by Berg and Duthie (1977). 

The most probable scenario, is that it consists of a {\bf triple system}, were BL Cam A (the SX Phe Star) is the primary with a mass of 0.99 $M_{\odot}$, BL Cam B should have a mass $m_{B} \ge 0.46 \ M_{\odot}$ a semi-major axis $a_B \ge$ 0.6 UA; BL Cam C should have a mass $m _{C} \ge 0.030 \ M_{\odot}  (i.e. \ge 31.5 \ M _{Jupiter})$ semi-major axis $ a _C \ge  4.4 \ UA.$

 \subsection{Photometry and Spectroscopy.}

As a part of an international campaign, we did simultanous photometric and spectroscopic observations at San Pedro M\'artir Observatory, on February and September, 2008. We used the 2.12m telescope with a Boller \& Chivens spectrograph and the 0.84 m telescope with CCD SITe1 photometer.  

Raw data spectrum of BL Cam obtained during September 24th, 2008 is shown. The changing behaviour of the flux of the star, following the star's periodicity is present. Vertical axis is the relative flux of the star for individual measurements (first panel).

Our February's 2008 observations give the following results. The gaussian fit of the equivalent width of the H$_\beta$ line of BL Cam (second panel), follows approximately the  photometric light curve (third panel), as can be seen from data obtained on February 29.  This Hydrogen line, does  not follows the {\it broad} photometric minimum observed.  The equivalent width curve, shows a phase lag of approximately 7 to 11 min  (i.e. 0.12 to 0.20) from the photometric period  P$_0$ = 56.3 min of the ligth curve of the star.


 \begin{figure}[!t]
  \includegraphics[width=\columnwidth]{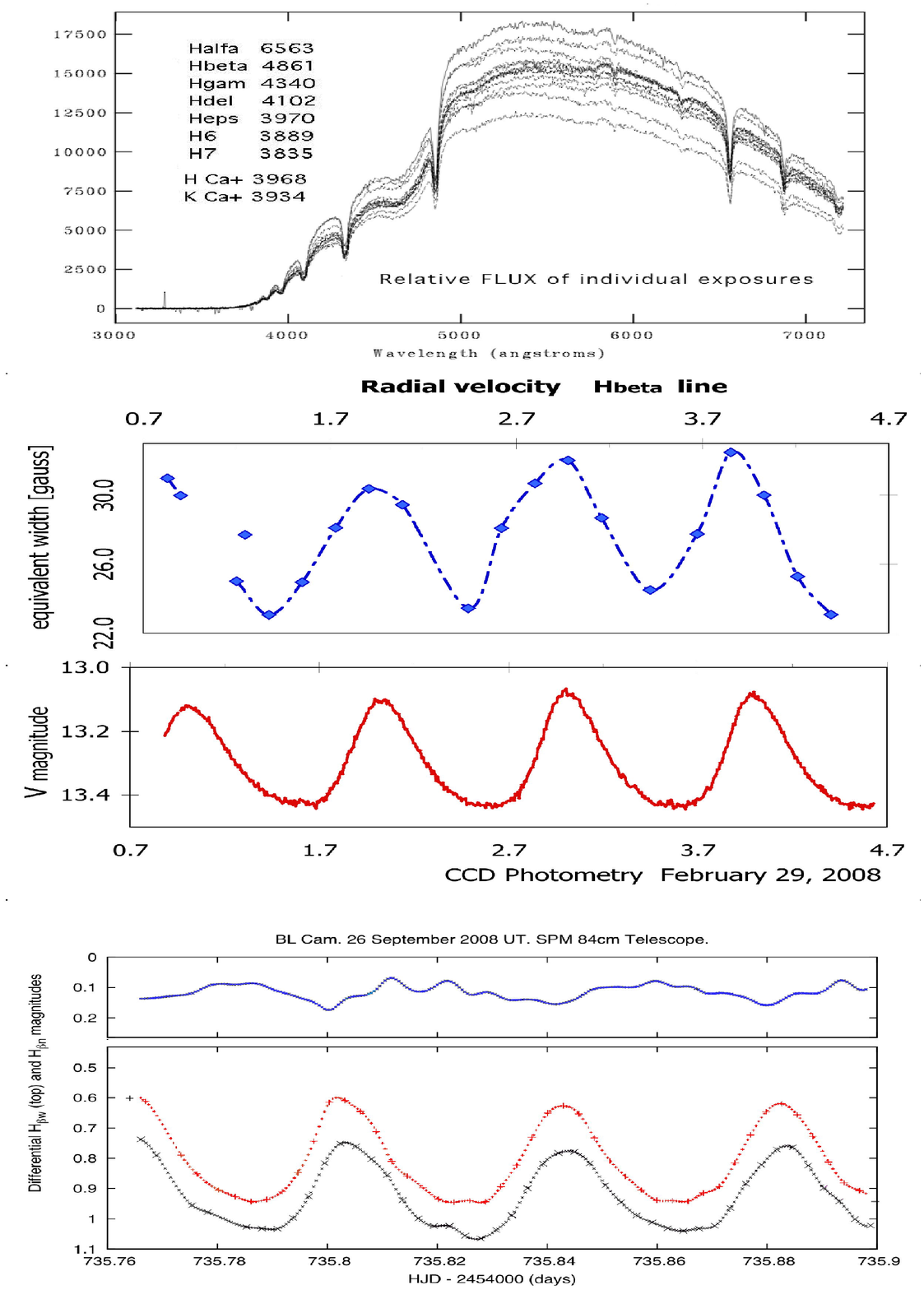} 
 \end{figure}

  \subsection{Narrow Band Photometric Observations.}

During September's run, the narrow {\it b}, {\it y} and H$_{\beta \ (n, w) } $ filters of the Str\"omgren's system, follow the similar behaviour of the very well studied Johnson's V filter, with a well pronounced maximum and a broad minimum (H$_{\beta \ (n, w) } $ on fifth panel). 

We fit the H$_{\beta \ (n, w)} \ $ values with a cubic spline function and computed the difference as shown (fourth panel).  It is clear that there is an interesting component that requires some physical explanation. Further analysis of these observations, will be published elsewhere.
 
{This research was done as part of PAPIIT IN108106 and IN114309 Program of UNAM.}


 \end{document}